\documentclass[11pt]{article}
\usepackage{amsmath}
\usepackage{amssymb}
\usepackage{amsthm}
\usepackage{amsfonts}
\usepackage{algorithm}
\usepackage{algorithmic}
\usepackage{graphicx}
\usepackage{pdfpages}
\usepackage{xcolor,colortbl}
\usepackage{multicol}
\usepackage{url}
\usepackage{subfigure}

\definecolor{Gray}{gray}{0.92}
\definecolor{LightCyan}{rgb}{0.88,1,1}

\newcolumntype{a}{>{\columncolor{Gray}}c}
\newcolumntype{b}{>{\columncolor{white}}c}

\usepackage{bbm,epsfig,graphics,epic,color,rotating,color}

    \usepackage[left=2.5cm,right=2.5cm,top=2.3cm,bottom=2.5cm]{geometry}

\numberwithin{equation}{section}
\def\E{{\mathbb E}}

\newcommand{\Z}{{\mathbb Z}}

\newtheorem*{theorem*}{Theorem}
\newtheorem{theorem}{Theorem}[section]
\newtheorem{lemma}[theorem]{Lemma}
\newtheorem{proposition}[theorem]{Proposition}

\newtheorem{conjecture}[theorem]{Conjecture}

\theoremstyle{definition}

\title{Lack of phase transitions in staggered magnetic systems. A comparison of uniqueness criteria.}
\author{Roberto Fern\'andez, Manuel Gonz\'alez-Navarrete, \\ Eugene Pechersky and Anatoly Yambartsev}

\date{}

\begin{document}

\maketitle

\vspace{2pt}

\begin{abstract}
We study a ferromagnetic Ising model with a staggered cell-board
magnetic field previously proposed for image processing [Maruani et al., Markov Processes Relat. Fields 1 (1995) \cite{MPS}]. We
complement previous results on the existence of phase transitions at
low temperature  [Gonz\'alez-Navarrete et al., J. Stat. Phys. 162 (2016)] by determining bounds to the region of
uniqueness of Gibbs measures.  We establish sufficient rigorous
uniqueness conditions derived from three different criteria: (1) Dobrushin criterion [Dobrushin, Theory Probab. Appl. 13 (1968) ], (2) Disagreement percolation [van den Berg and Maes, Ann. Probab. 22 (1994)] and (3) Dobrushin-Shlosman criteria [Dobrushin and Shlosman,  in Statistical Physics and Dynamical Systems. Rigorous Results. (1985)]. These conditions are subsequently solved
numerically and the resulting uniqueness regions compared.
\end{abstract}

\section{Introduction}
\label{sec:intro}

In his seminal work \cite{Dbr1}, Dobrushin introduced a constructive
sufficient condition for absence of phase transitions in statistical
mechanics.  The condition is ``constructive" in the sense that its
validity can be determined by a finite (usually small) number of
computations.  This feature opened the way to  computer-assisted
proofs of uniqueness. Dobrushin criterion was later generalized in
two directions. On the one hand, Dobrushin and Pecherski \cite{DP} and
Dobrushin and Shlosman \cite{DS} produced constructive generalizations
that have been extensively used to obtain, or refine, a number of
uniqueness results in classical statistical mechanics (see also \cite{We}). On the other
hand, van den Berg \cite{vdB} introduced the alternative approach
called \textit{disagreement percolation}, later improved by van den
Berg and Maes \cite{vdBM}.  The approach is based in comparing
probabilities of disagreements between two realizations of the model
with a corresponding model of independent percolation.

These three criteria ---Dobrushin (DC), Dobrushin-Shlosman (DS)
and disagreement percolation (DP)--- have different optimal domains of
application, and they may produce complementary results for the same
model.  In this paper we apply them to the cell-board Ising models
introduced in \cite{MPS} in reference to image processing.  The case
of 1x1 cells corresponds to the antiferromagnetic Ising model which
has been an important laboratory for the different uniqueness
criteria.  Indeed, the model was first studied via DS in \cite{DKS} in
the vicinity of the critical external field, namely $h_c=4J$ (see
\eqref{crit} below). Later, van den Berg used this model to
introduce the DP method \cite{vdB}.

Other than the 1x1 case, published papers on cell-board focus in the
phase-transition region. This region was proven to be non-trivial in
\cite{GPY}, using a Peierls-type argument based on a chessboard
inequality obtained via reflection positivity (see \cite{Bi,Shl}). The present paper is the first one, to our knowledge, establishing uniqueness regions for the Ising model with cell-board external field with cells of size $L_1\times L_2$. However, it is worth mentioning the alternated external field in Nardi et al.~\cite{NOZ} (it corresponds to the case $L_1=\infty$ and $L_2=1$), they provide a phase transition and obtained uniqueness regions by applying cluster expansion technique \cite{MaMi} through a translation-invariant property of clusters. The cluster expansion is a large area and here we limit to the non-cluster expansion criteria. 


%
We conclude that the best criterion to deal with the class of cell-board Ising models is the DS criterion. Essentially, DC criterion is not able to identify the influence of parameters $L_1$ and $L_2$. In the case of DP using independent site percolation, we have the same limitation.
Although it is possible to observe that these latter give us complementary results depending on the strength of the external field. In addition, the numerical results allow us to conclude that uniqueness holds for all temperature whenever the external field is larger than a critical value (see Conjecture \ref{uniqueness}). The main difficulty is the computational cost of such calculations. We remark that rigorous proof of this fact is still an open problem.

\section{The uniqueness regions of cell-board Ising models}
\label{sec:model}

\subsection{Setup and overview}\label{ssec:setup}

Let $\Omega = \{-1,+1\}^{\Z^2}$ be a set of all spin configurations on $\Z^2$. For any $\sigma \in \Omega$ the formal Hamiltonian is defined  as
\begin{equation}
\label{ham.chess}
H(\sigma)= -J \displaystyle\sum_{\langle t, s \rangle} \sigma(t) \sigma(s) - \displaystyle\sum_s h(s) \sigma(s)\;,
\end{equation}
where $J > 0$ is the ferromagnetic interaction constant, $\langle t, s \rangle$ denotes unordered pairs of nearest neighbours $s, t \in \mathbb{Z}^2$ and the function $h(s)$ represents periodical cell-board external fields, defined as follows. For each pair $n, m$ of integers we associate the cell 
\begin{equation}
C(n,m) =  \bigl\{ (t_1,t_2) \in \Z^2 :  nL_1 \le t_1 < (n+1)L_1\;,\; mL_2 \le t_2 < (m+1)L_2 \bigr\},
\end{equation}
where $L_1$ and $L_2$ are given positive integers, representing size of cells, and let 
\begin{equation}
\mathbf{Z}_+ = \bigcup_{\scriptstyle n,m:\atop  \scriptstyle n+m \text{ is even}} C(n,m) 
\quad,\quad \mathbf{Z}_- = \Z^2 \setminus \mathbf{Z}_+\;.
\end{equation}
We interpret $\mathbf{Z}_+$ as the set of white cells and $\mathbf{Z}_-$ as the set of black cells of the infinite ``chess-board'' $\mathbb{Z}^2$.  See Figure \ref{figTh} for particular cases $L_1=3, L_2=2$ and $L_1=L_2=2$.
Then, for a fixed $h\ge 0$ we define the configuration of external field $(h(s), s\in \mathbb{Z}^2)$, where
\begin{equation}
\label{external}
h(s)=
\left\{
\begin{array}{rl}
h,&\text{ if }s\in \mathbf{Z}_+,\\
-h,&\text{ if }s\in \mathbf{Z}_-\;.
\end{array}
\right.
\end{equation}
Note that cell-board external field \eqref{external} may create a non-constant ground state $\sigma_c$, which we call cell-board configuration, where $\sigma_{c}(t)=+1$, for all $t\in \mathbf{Z}_+$ and $\sigma_{c}(t)=-1$, whenever $t\in \mathbf{Z}_-$.  These ground states appear, however, at precisely tuned values.   Indeed, the phase diagram at zero temperature, that is $T=0$, is given by the following result from \cite{MPS}.

\begin{theorem}(Maruani et al. \cite{MPS})
\label{pecherski}
If
\begin{equation}\label{crit}
 h < h_c:= \frac{2J}{L_1} + \frac{2J}{L_2}\;,
 \end{equation}
 then there exist two periodical ground states, namely the constant configurations $\sigma^+ \equiv +1$ and $\sigma^- \equiv -1$. If $h > h_c$, then $\sigma_{c}$ is the unique periodical ground state.
\end{theorem}

In the critical case $h=h_c$, there exist infinitely many ground states.
\smallskip

The coexistence of ground states for weak fields was shown in \cite{GPY} to extend to a coexistence of Gibbs states for low enough temperatures. Let, as usual, $\beta=1/T$ be the inverse temperature, then Theorem 2 from \cite{GPY} states

\begin{theorem}(Gonz\'alez-Navarrete et al. \cite{GPY})
\label{phase transition}
Consider the cell-board model on $\mathbb{Z}^2$ defined by the Hamiltonian \eqref{ham.chess}. If \eqref{crit} holds, then there exists $\beta_c : = \beta_c (J,h,L_1,L_2)$, such that for any $\beta > \beta_c$, there exist two distinct measures $\mu^+_{\beta}, \mu^-_{\beta} \in \mathcal{G}_{\beta}$, which satisfy $\mu^{\pm}_{\beta} (\sigma(t) = \pm 1 ) > \frac{1}{2}$ for all $t\in\mathbb{Z}^2$. 
\end{theorem}

The region $h \ge h_c$ has not been studied so far, and it is in fact the object of the present work.  The underlying conjecture is the uniqueness of the Gibbs measure for all temperatures. 
\begin{conjecture}\label{conj:r1}
\label{uniqueness}
For the cell-board Ising model on $\mathbb{Z}^2$, as defined in \eqref{ham.chess},  there exists a unique Gibbs measure for all temperature, whenever $h\ge h_c$ .
\end{conjecture}

It is worth mentioning that the antiferromagnetic Ising model with uniform external field corresponds to the case $L_1=L_2=1$, see Corollary 1 from \cite{GPY}.  For this case, the existence of multiple Gibbs states at low temperatures was proved by Dobrushin \cite{Dbr}, see also Fr\"olich et al.\ \cite{Fro}.  In the opposite direction, the uniqueness criteria of Dobrushin and Shlosman \cite{DS} and of van den Berg \cite{vdB} prove Conjecture \ref{uniqueness} for 
\begin{equation}\label{eq:r01}
h\;>\;h_c \;=\; 4J,
\end{equation}
in agreement with \eqref{crit}.

In addition, the model considered by Nardi, Olivieri and Zahradn\'i{}k \cite{NOZ} can be compared with a cell-board Ising, by letting $L_1=\infty$ and $L_2=1$, in that work it was proven uniqueness at low-temperature for $h> 2J$, by using the method of cluster expansion (see Sections 2.1 and 2.2 in \cite{NOZ}). Note that Conjecture \ref{uniqueness} completes the results from \cite{NOZ} for all temperature. However, it is worth mentioning that Nardi et al. \cite{NOZ} considered different intensities ($h_1$ and $h_2$) on $\mathbf{Z}_-$ and $\mathbf{Z}_+$, in the external field \eqref{external}. They also proved the phase transition on $h_1=h_2< 2J$ for low enough temperature (see Theorem 3.1 in \cite{NOZ}). Further, we refer to that model as NOZ Ising model.

\subsection{Results}

We apply three uniqueness criteria: Disagreement percolation (DP),  Dobrushin criterion (DC) and Dobrushin-Shlosman criterion (DS).  We obtain rigorous bounds on the corresponding uniqueness regions, involving inequalities that are subsequently compared numerically.  Unless stated otherwise, the results below apply to cell-board Ising models on $\mathbb{Z}^2$ with general $L_1,L_2\ge 1$.
Let denote
\begin{equation}\label{eq:r1}
p(J/T, h/T)\;:=\; \frac{\sinh ( 8J / T ) }{ \cosh( 2h/T) + \cosh( 8J/T )} \;.
\end{equation}
and $p_c$ being the critical percolation probability of the independent site percolation in $\mathbb{Z}^2$. 

 \begin{proposition}[\bf Disagreement percolation criterion]
\label{PropoDP1}

The cell-board Ising model has a unique Gibbs measure whenever

\begin{equation}
\label{T0DP}
T > T^{DP}(J,h) \equiv  T^{DP}(J,h,L_1,L_2):= \inf \left\{T \ge 0 : p(J/T, h/T)\le p_c \right\}.
\end{equation}
In particular, there exists a unique Gibbs measure for all temperatures, whenever $h\ge 4J$.
\end{proposition}

In the next result we use the notation
 \begin{equation}\label{eq:r3}
 \gamma(J/T,h/T)\;:=\;  \max_{n \in \{-3,-1,1,3\}} \frac{\sinh( 2 J/T) }{\cosh(2 (J n+ h)/T) + \cosh(2 J/T ) }\; .
 \end{equation}
 \begin{proposition}[\bf Dobrushin criterion]
 \label{PropoDo}

 The cell-board Ising model has a unique Gibbs measure whenever $\gamma(J/T,h/T) < 1/4$, which define the DC-Temperature
 \begin{equation}\label{T0DC}
 T^{DC} (J,h) \equiv T^{DC}(J,h,L_1,L_2):= \inf \bigl\{ T \ge 0: \gamma(J/T,h/T) = 1/4 \bigr\}\;,
  \end{equation}
Moreover, $T^{DC} (J,h)$ satisfies:
 
 \begin{itemize}
 \item[(i)] $T^{DC}(J,h)$ is symmetric around $2J$ on the interval $h\in [0,4J]$, and it is symmetric around $J$ ($3J$) on the interval $h\in [0,2J]$ ($h\in [2J,4J]$), that is,
\begin{equation}
\label{Dob0}
T^{DC}(J,h) =T^{DC}(J,h +2J) =T^{DC}(J,2J-h) = T^{DC}(J,4J-h),
\end{equation}
for all $h\in [0,2J]$.

\item[(ii)] The following inequalities hold
 \begin{equation}
\label{Dob1}
\frac{4J}{\ln 3}  \le T^{DC}(J,h) \le \frac{2J}{\ln(5/3)}\;,\  \mbox{ when }h\in [0,4J].
\end{equation}
 \end{itemize}
 \end{proposition}
 
 Since both criteria use single-site distributions, the estimates $T^{DP}$ in \eqref{T0DP} and $T^{DC}$ in \eqref{T0DC} do not depend on $L_1$ and $L_2$ values. Then, the bounds are the same as for the antiferromagnetic Ising with uniform external field. The details are included in Section \ref{sec:uniq}.
 
 \begin{proposition}[\bf  Dobrushin-Shlosman criterion]\label{prop:r10}
 There exist a sequence of parameters 
 \begin{equation}\label{eq:r10}
\gamma_n(J/T,h/T,L_1,L_2),\quad n=1,2,\ldots\,,
 \end{equation}
(will be defined in \eqref{DSgamma} and \eqref{alpha}) such that the cell-board Ising model has a unique Gibbs measure whenever 
\begin{equation}\label{eq:r4B}
\gamma_n(J/T,h/T,L_1,L_2)\;<\: 1\quad \mbox{for some } n\in \mathbb{N}\;.
 \end{equation}
 Furthermore, the DS-critical lines
\begin{equation}
\label{hDS1}
 l^{DS}(n;J,L_1,L_2) =\bigl\{(h,T) \in \mathbb{R}_+^2 : \gamma_n(J/T,h/T,L_1,L_2) = 1 \bigr\}
\end{equation}
 and the uniqueness regions
 \begin{equation}
\label{LDS1}
 L^{DS}(n;J,L_1,L_2) =\bigl\{(h,T) \in \mathbb{R}_+^2 : \gamma_n(J/T,h/T,L_1,L_2) < 1 \bigr\}
\end{equation}
 satisfy the following relations for all $J>0$:
 
\begin{itemize}
 \item[(i)] For any $L_1$ and $L_2$ finite, such that $\min\{L_1,L_2\}\ge n$,
\begin{equation}
\label{DS1}
l^{DS}(n;J,L_1,L_2) = l^{DS}(n;J,n,n)\;.
\end{equation}

\item[(ii)] For any $n\ge 2$,
\begin{equation}
\label{DS2}
L^{DS}(n;J,n-1,n-1)\subseteq L^{DS}(n;J,n,n).
\end{equation}

\item[(iii)] For any $L_1,L_2 < \infty$,
\begin{equation} \label{eq:r15}
l^{DS}(2;J,L_1,L_2) =l^{DS}(2;J,1,1), \ \text{ for all }  J > 0.
\end{equation}
\end{itemize}

\end{proposition}

The expression for the functions $\gamma_n$ is detailed in Section \ref{sec:DS}. In particular $\gamma_1=4 \gamma$, the latter being the function for the Dobrushin criterion in \eqref{eq:r3}.  In addition, note that for fixed $n$, item \textit{(i)} implies that if the cell size grows larger than $n$, then $l^{DS}$ estimates stay constants. In the sense of \textit{(ii)}, consider cells as squares, then the $l^{DS}$ estimates obtained by $n$ for a cell of side $n$ gives the same or worse estimate than a smaller cell. Finally, in \textit{(iii)} we fix $n=2$, then $l^{DS}$ does not change for any finite $L_1 \times L_2$ cell.

\bigskip

\begin{figure}[ht]
\begin{center}
\includegraphics[scale=0.37]{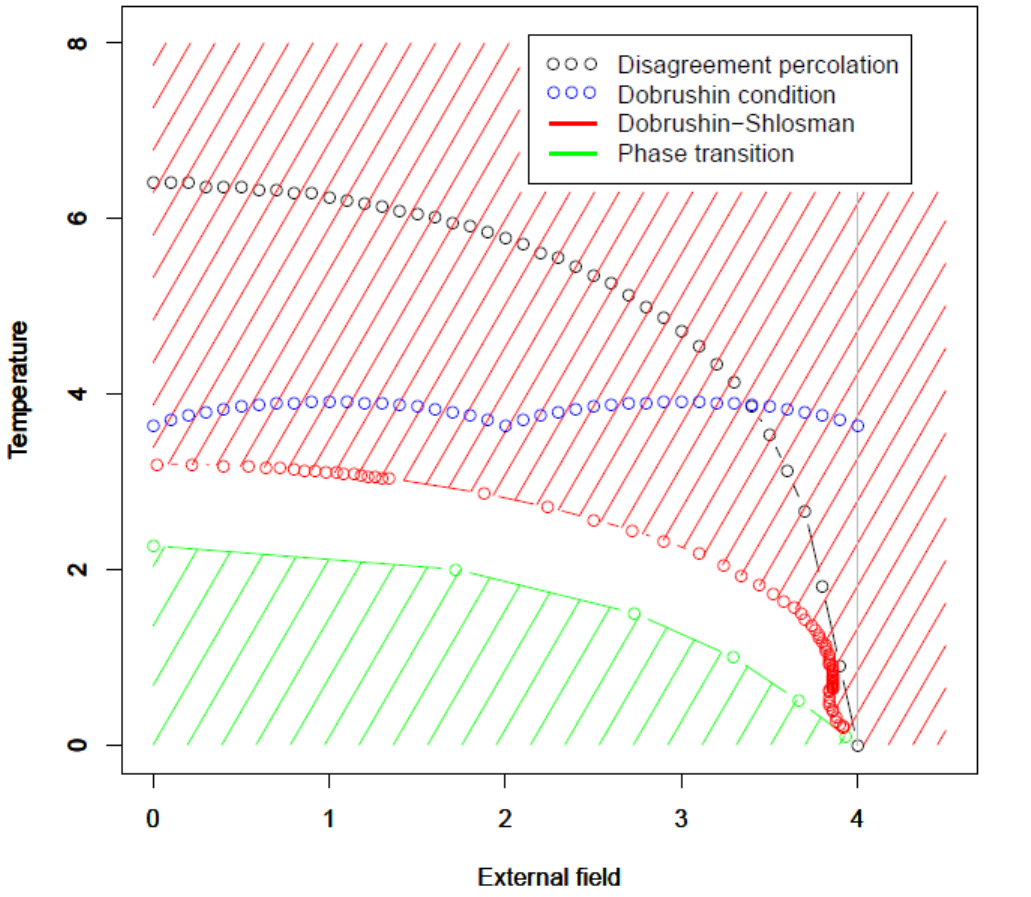}
\end{center}
\caption{{\scriptsize Bounds for the case $L_1=L_2=1$. Phase transition as obtained in \cite{LD}. The region of uniqueness was obtained with Dobrushin-Shlosman criterion by square of size $n=3$. The critical $h_c=4$.}}\label{fig:Anti}
\end{figure}

\begin{figure}[ht]
\begin{center}
\subfigure[]{
\includegraphics[scale=0.35]{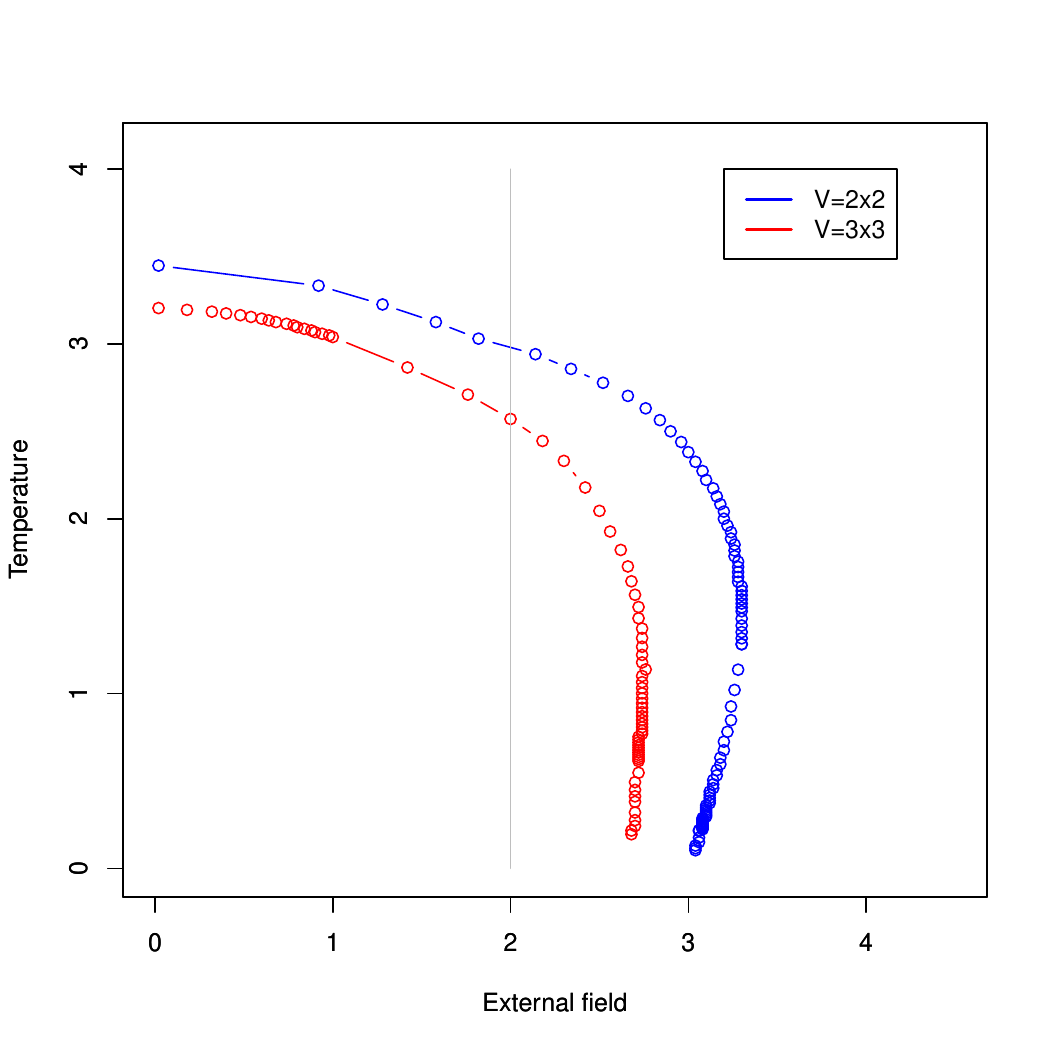}
}
\hspace{0.2cm}
\subfigure[]{\includegraphics[scale=0.35]{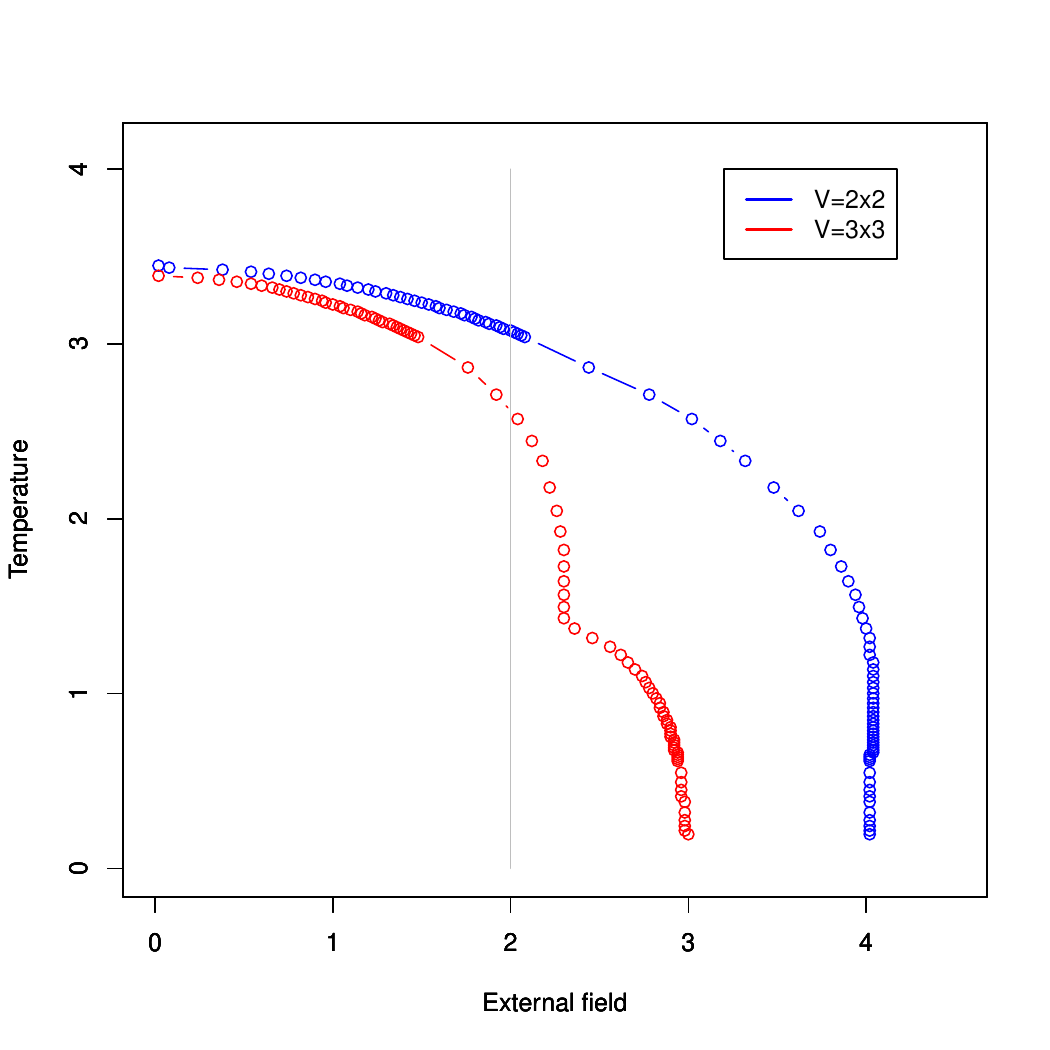}
}
\end{center}
\caption{{\scriptsize Dobrushin-Shlosman estimates, taking squares $n=2$ and $n=3$. (a) The Nardi-Olivieri-Zahradn\'ik Ising model \cite{NO,NOZ}. (b) The case $L_1=L_2=2$. For both models, the vertical line is critical value $h_c=2$.}}\label{fig:V2x2}
\end{figure}

\begin{figure}[h!]
\begin{center}
\includegraphics[scale=0.36]{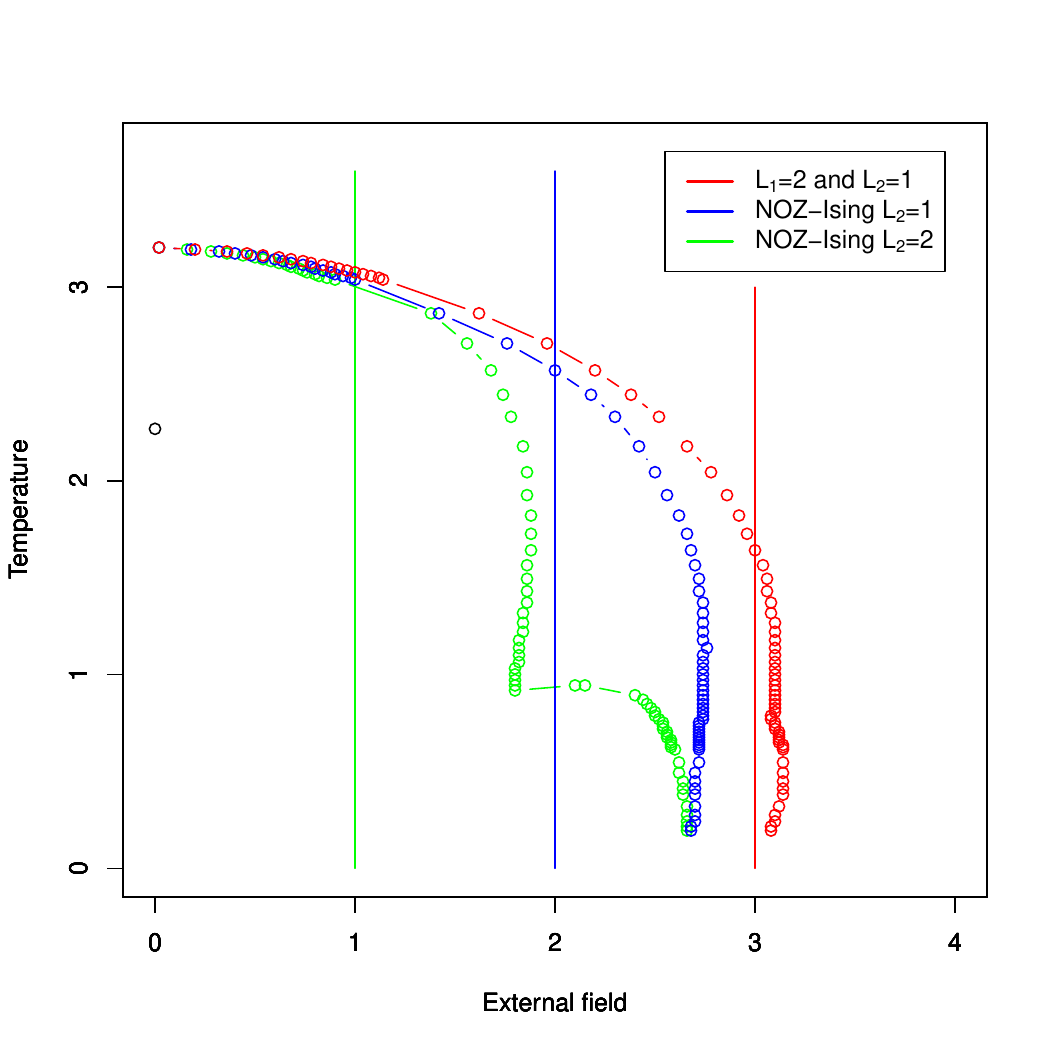}
\end{center}
\caption{{\scriptsize Dobrushin-Shlosman estimatives with square size $n=3$. The case $L_1=2$ and $L_2=1$, critical value $h_c=3$. NOZ Ising models, in case $L_2=1$ with $h_c=2$, and $L_2=2$ with $h_c=1$. Vertical lines represent critical values of corresponding model.}}\label{fig:L2x1}
\end{figure}

In Figures \ref{fig:Anti}-\ref{fig:L2x1} we summarize numerical comparisons of the above criteria.  Without loss of generality we set $J=1$ that is, we plot in terms of scaled parameters $h/J$ and $T/J$.
\smallskip

Figure \ref{fig:Anti} shows a comparison of the three criteria for the case $L_1=L_2=1$, namely the antiferromagnetic Ising model with uniform external field.  The figure shows the 
curves of $T^{DP}(1,h)$, defined by \eqref{T0DP}, $T^{DC}(1,h)$ defined by \eqref{T0DC} and $l^{DS}(3;1,T,1,1)$ defined by \eqref{hDS1}. In addition it shows the limit phase-transition curve obtained from \cite{LD} and references therein.

It is apparent that DS criterion allows better estimates in this particular external field, both for small and large values of $h$. In particular it offers another proof of Conjecture \ref{conj:r1} with $h_c=4J$.  The DC criterion, in turns, is better than DP for low values of the magnetic field and inversely for values close to the critical field.   This agrees with the observation \cite{vdBM} that DC criterion is better than DP for ferromagnetic models, while DP is better for antiferromagnetic models.  
\medskip

Figures \ref{fig:V2x2} and \ref{fig:L2x1} present DS lines for different values of $L_1$ and $L_2$.  
Figure \ref{fig:V2x2} compares estimates for cell-board Ising with $L_1=L_2=2$ and the NOZ Ising model \cite{NO,NOZ}, that is, $L_1=\infty$ and $L_2=1$, for square sizes $n=2$ and $n=3$. We remark that both models have the same critical value $h_c=2$. Although the DS estimates do not reach the critical line, it is expected that for large squares it will be done, which complement the uniqueness obtained for low temperature in \cite{NOZ}.

Figure \ref{fig:L2x1} presents three estimates obtained for $n=3$. The one for the model $L_1=2$ and $L_2=1$ almost reaches the critical value $h_c=3$. However, it is possible to prove that this estimate is the same as long as $L_2=1$ and $2\le L_1< \infty $:
\begin{equation}
\label{L21}
l^{DS}(3;J,L_1,1) = l^{DS}(3;J,2,1), \ \text{ for all finite } \ L_1 \ge 2.
\end{equation}

In fact it also holds true that
\begin{equation}
l^{DS}(3;J,L_1,2) = l^{DS}(3;J,2,2), \ \text{ for all finite } \ L_1 \ge 2.
\end{equation}

The DS estimate for the NOZ Ising model exceeds the critical $h_c=2$. We also analyse the external fields proposed in Section 5 of \cite{GPY}, namely $L_1=\infty$ and $L_2\ge 2$, which generalize the Ising model with an alternating stripes external field studied in \cite{NO,NOZ}. The DS estimates with $n=3$ is the same for all cases, that is, the estimate in Figure \ref{fig:L2x1} remains constant for any $L_2 \ge 2$.

\section{Proofs}
\label{sec:uniq}

First of all, we define the Wasserstein (Kantorovich) and total variation distances (for details, see \cite{GS}). For a finite region $\Lambda\subset \mathbb{Z}^2$ and configurations in $\Omega_\Lambda=\{-1,1\}^\Lambda$, let $\rho_\Lambda(\cdot, \cdot )$ a metric on $\Omega_\Lambda$, Gibbs measures $\mu_1, \mu_2$ and joint probability $\nu$ on $\Omega_\Lambda \times \Omega_\Lambda$ such that $$ \nu ( B \times \Omega_\Lambda ) = \mu_1 (B) \ \ \text{ and } \ \ \nu(\Omega_\Lambda \times B) = \mu_2(B), \ \ B \subset \Omega_\Lambda,$$
we denote $K(\mu_1, \mu_2)$ the set of all that joint measures $\nu$. The Wasserstein distance is defined by

\begin{equation}\label{dW}
d_W(\rho_\Lambda;\mu_1,\mu_2) = \inf_{\nu \in K(\mu_1,\mu_2)} \left\{ \int_{\Omega_\Lambda \times \Omega_\Lambda} \rho_\Lambda(\sigma, \sigma^{\prime} ) \nu(d\sigma, d \sigma^{\prime})\right\} 
= \inf_{ \nu \in K(\mu_1,\mu_2)} \left\{ \E_{\nu} ( \rho_\Lambda(\sigma, \sigma^{\prime} ) )\right\}.
\end{equation}

Moreover, the total variation distance is given by

\begin{equation}
d_{TV}(\mu_1,\mu_2) = \sup_{A \subset \Omega_\Lambda} | \mu_1(A) - \mu_2(A)|,
\end{equation}
an alternative form was given in \cite{Li}
\begin{equation}
\label{dTV}
d_{TV}(\mu_1,\mu_2)= \inf_{ \nu \in K(\mu_1,\mu_2)} \left\{ \nu( (\sigma, \sigma^{\prime} ): \sigma \neq \sigma^{\prime}) \right\}.
\end{equation}

In particular, disagreement percolation (DP) and Dobrushin (DC) criteria are based on the total variation distance between finite-region Boltzmann-Gibbs distributions for different boundary conditions. Let a configuration $\sigma_{\Lambda} \in \Omega_{\Lambda}$, the weights for the Boltzmann-Gibbs distribution for a region $\Lambda\subset \mathbb{Z}^2$ with external configuration $\omega$ at inverse temperature $\beta$ take the form
\begin{equation}
\label{eq:r20}
\mu_{\beta,\Lambda}(\sigma_\Lambda\mid\omega)\;=\; 
\frac{e^{-\beta H_\Lambda(\sigma_\Lambda\mid\omega)}}{\sum_{\xi_\Lambda}e^{-\beta H_\Lambda(\xi_\Lambda\mid\omega)}},
\end{equation}
where $H_\Lambda(\cdot \mid\omega)$ is the Hamiltonian for configurations on $\Omega_\Lambda$ with boundary condition $\omega$.
The total variation distance between two distributions with external conditions $\omega$ and $\omega'$ is defined by
\begin{equation}
\label{eq:r21}
d_{TV}\bigl(\mu_{\beta,\Lambda}(\cdot\mid\omega), \mu_{\beta,\Lambda}(\cdot\mid\omega')\bigr)
\;=\; \frac12 \sum_{\sigma_\Lambda}\bigl| \mu_{\beta,\Lambda}(\sigma_\Lambda\mid\omega)
- \mu_{\beta,\Lambda}(\sigma_\Lambda\mid\omega')\bigr|\;.
\end{equation}
More specifically, DP and DC criteria depend on the distance between single-site distributions. In this case the distance admits the simpler form
\begin{equation}
\label{dmu}
d_{TV}(\mu_{\beta,s}(\cdot | \omega),\mu_{\beta,s}(\cdot | \omega^{\prime})) 
= \bigl| \mu_{\beta,s}(+1 | \omega) - \mu_{\beta,s}(+1 | \omega^{\prime}) \bigr|
\end{equation}
[we have denoted $\mu_{\beta,s}:= \mu_{\beta,\{s\}}$].

In the case of Dobrushin-Shlosman criterion (DS), we remark that in their seminal work \cite{DS} it was used the Wasserstein distance \eqref{dW} to construct a condition stating that: if for a finite $\Lambda \subset \Z^2$, the Wasserstein distance between two distributions with external conditions $\eta$ and $\eta'$, respectively, is bounded in a suitable way, then there is a unique Gibbs measure. In the literature, the control of Wasserstein distance has been also used to obtain convergence results in long-term behaviours of Markov chains (see, for instance \cite{AH,MS,RS}).

Now, let denote $\partial \Lambda = \{  t \in \Z^2\setminus \Lambda :  \langle t, s \rangle, \text{ for some } s \in  \Lambda\}$, as explained, by $\langle t, s \rangle$ we mean that $t$ and $s \in \Z^2$ are nearest neighbours. Then, the criterion can be stated as follows,

\begin{theorem}
\label{DSthm}
Suppose for a given Gibbs measure $\mu$, there exists a volume $\Lambda$, such that: there exists a function $\alpha_t \ge 0$, $t \in \partial \Lambda$ with the following properties
\begin{enumerate}
\item For any $ t \in \partial \Lambda$ and any $\eta, \eta^{\prime} \in \Omega, \eta(s) = \eta^{\prime} (s)$ for $s\neq t$
\begin{equation}
\label{ThdW}
 d_W(\rho_\Lambda;\mu_\Lambda(\cdot| \eta) ,\mu_\Lambda(\cdot| \eta^{\prime})) \le \alpha_t \cdot \rho_t(\eta(t),\eta^{\prime}(t)).
\end{equation}
where $d_W$ is the Wasserstein distance \eqref{dW} and
\begin{equation}
\label{Thrho}
\rho_\Lambda(\sigma, \sigma^{\prime}) = \displaystyle\sum_{t \in \Lambda} \rho_t(\sigma(t),\sigma^{\prime}(t)),
\end{equation}
for $|\Lambda| < \infty$, where $\rho_t$ is a metric on the spin space.
\item \begin{equation}
\label{Thgam}
\displaystyle\frac{1}{|\Lambda|}\sum_{t \in \partial \Lambda} \alpha_t = \gamma < 1.
\end{equation}
Then, there exists a unique Gibbs measure.
\end{enumerate}
\end{theorem}

Now, we are in condition to state the proofs of the main results.

\subsection{Proof of Proposition \ref{PropoDP1}}

The disagreement percolation introduced in the DP criterion \cite{vdB} is an independent site percolation model in which each site $s \in \mathbb{Z}^2$ is open with probability 
\begin{equation}
\label{pis}
p_s(\beta,J,h(s)) = \displaystyle\max_{\omega, \omega^{\prime} } d(\mu_{\beta,s}(\cdot | \omega),\mu_{\beta,s}(\cdot | \omega^{\prime}))\;.
\end{equation}

The criterion states that there exists a unique Gibbs measure 
in the absence of percolation, i.e.\ if
\begin{equation}\label{perc.def}
P (\text{there exists an infinite open path}) = 0
\end{equation}
where $P$ is the product probability defined by \eqref{pis}.   In particular this happens if
\begin{equation}
\label{uniccond1}
p := \displaystyle\sup_{s \in \Z^2} p_s(\beta,J,h(s)) < p_c (\Z^2),
\end{equation}
where $p_c (\Z^2)$ is the critical probability of the independent site percolation model on the square lattice. Sufficient conditions can be obtained using lower bounds on $p_c$.  The bound $p_c>1/2$ in \cite{Hi} leads to the proof of uniqueness for cell-board Ising models with $h>1/4$ (see below).  The lower bound $p_c>0.556$ in \cite{vdBE} was used to generate the corresponding curve in Figure \ref{fig:Anti}.

To determine $p$ in \eqref{uniccond1} we consider the set of nearest-neighbors of $s$
\begin{equation}
    \label{neigh}
    \mathcal N(s) = \{ t \in \Z^2 : |t-s|=1  \}
\end{equation}
and introduce
\begin{equation*}
n = \displaystyle\sum_{j \in \mathcal{N}(s)} \omega(j) \ \ \text{and} \ \ n^{\prime} = \displaystyle\sum_{j \in \mathcal{N}(s)} \omega^{\prime}(j),
\end{equation*}
where $n,n' \in \{ -4, -2, 0, 2, 4 \}$.
Then
\begin{equation*}
\begin{aligned}
p & = \max_s \max_{n, n^{\prime} }
\Bigl| \frac{1}{1 + \exp(-2\beta (J n + h(s)))} - 
\frac{1}{1 + \exp(-2\beta (J n' + h(s))) } \Bigr|  \\
& =  \max_s \Bigl( \frac{1}{1 + \exp(-2\beta (4 J + h(s)))}
- \frac{1}{1 + \exp(-2\beta (-4J + h(s)))} \Bigr) \\
& =  \frac{1}{1 + \exp(-2\beta (4 J + h))}
- \frac{1}{1 + \exp(-2\beta (-4J + h))} \;.
\end{aligned}
\end{equation*}
The last equality is due to the fact that the argument of the max is invariant under the change $h(s)\to -h(s)$.

Simple manipulations yield
\begin{equation}\label{11}
p = \frac{\sinh ( 8\beta J ) }{ \cosh(2\beta h) + \cosh( 8\beta J)}\;.
\end{equation}
\smallskip

The last statement of the proposition follows from the inequalities
\begin{equation}\label{eq:r30}
p = \frac{\sinh ( 8\beta J ) }{ \cosh(2\beta h) + \cosh( 8\beta J)} \le 
\frac{\sinh ( 8\beta J ) }{ 2\cosh( 8\beta J)} = \frac{1}{2} \mbox{tanh} ( 8\beta J) <  \frac{1}{2}
< p_c\;.
\end{equation}

\subsection{Proof of Proposition \ref{PropoDo}}

Dobrushin criterion \cite{Dbr1} establishes that there is a unique Gibbs state if
\begin{equation}
\label{Docri}
\sup_{s \in \Z^2} \displaystyle\sum_{t \in  \mathcal{N}(s) }\max_{\omega, \omega^{(t)} \in \Omega } d_{TV}(\mu_{\beta,s}(\cdot | \omega),\mu_{\beta,s}(\cdot | \omega^{(t)})) < 1.
\end{equation}
\noindent
where $\mathcal{N}(s)$ as defined in \eqref{neigh} and
\begin{equation}
\label{flip}
 \omega^{(t)}(s)=
\left\{
\begin{array}{rl}
\omega(s),&\text{for all }s \neq t,\\
-\omega(s),&s=t,
\end{array}
\right.
\end{equation}
which means that both configurations coincide at all sites different from $t$.  For cell-board Ising models this condition is equivalent to
\begin{equation}
\label{criti2}
\sup_{s \in \Z^2} \max_{\omega, \omega^{(t)} \in \Omega } d_{TV}\bigl(\mu_{\beta,s}(\cdot | \omega),\mu_{\beta,s}(\cdot | \omega^{(t)})\bigr) < \frac{1}{4}.
\end{equation}
Denoting
\begin{equation}
\label{eq:r40}
n = \displaystyle\sum_{j \in \mathcal{N}(s)\setminus t} \omega(j) = \displaystyle\sum_{j \in \mathcal{N}(s)\setminus t} \omega^{(t)}(j)\; \in\; \mathcal{M} =\{-3,-1,1,3\}\;,
\end{equation}
using \eqref{dmu}, then condition \eqref{criti2} becomes
\begin{eqnarray*}
\frac14 &>&\sup_{s \in \Z^2} \,\max_{\scriptstyle n\in \mathcal M \atop\scriptstyle  \omega(t)=\pm 1}\, \Big| \frac{\exp( \beta J (n+\omega(t)) + \beta h(s))}{2\cosh(\beta J (n+\omega(t)) + \beta h(s))} - \frac{\exp( \beta J (n-\omega(t)) + \beta h(s))}{2\cosh(\beta J (n-\omega(t)) + \beta h(s))} \Big| \\
&=&
\sup_{s \in \Z^2} \,\max_{\scriptstyle n\in \mathcal M \atop\scriptstyle  \omega(t)=\pm 1} \, \Big| \frac{\sinh( 2 \beta J \omega(t)) }{\cosh(2\beta (J n+ h(s))) + \cosh(2\beta J \omega(t)) } \Big| \\
&=&
\max_{\scriptstyle n\in \mathcal M } \,  \frac{\big|\sinh( 2 J/T )\big| }{\cosh(2 (J n+ h)/T) + \cosh(2 J /T) } \;.
\end{eqnarray*}
the last equality follows from the fact that the denominator in previous line is always positive, and absolute value of the numerator is invariant under changes in the signs of $\omega(t)$. Moreover, given that the function $\cosh(\cdot)$ is symmetric, the $\max$ on $n \in \mathcal{M}$ is invariant under changes in the signs of $h(s)$.
\smallskip

To see the symmetry properties \eqref{Dob0}, write the criterion \eqref{eq:r3} as
\begin{equation}\label{eq:r45}
\widetilde f(J/T,h/T)\;=\;\max_{n \in \mathcal{M} } f_n(J/T,h/T)  \;<\; \frac{1}{4}
\end{equation}
with\begin{equation}\label{DCcritfunc1}
f_n(J/T,h/T) :=  \frac{\sinh( 2  J/T) }{\cosh(2 (J n+ h)/T) + \cosh(2 J/T ) }\;,
\end{equation}
for $J>0$. Then, we have
\begin{equation}\label{DCcritfunc2}
\widetilde{f}(J/T,h/T) = \left\{ \begin{array}{ll} f_{-1} (J/T,h/T), & \text{ if } h \in [0,2J], \\  
f_{-3} (J/T,h/T), & \text{ if } h \in [2J, 4J]. \end{array} \right.
\end{equation}
All the symmetry properties of $T^{DC}(J,h)$ stated in \textit{(i)} are an immediate consequence of the symmetry properties of function $\widetilde{f}(J/T,h/T)$ for $h\in [0,4J]$.
\smallskip 

To prove the bounds \eqref{Dob1}, note that the extremal values of the function $T^{DC}(J,h)$ for $h$ on the interval $[0,4J]$ coincide with those attained on the interval $[0,J]$. They can, therefore, be determined studying $f_{-1}(J/T,h/T)$.  This function is monotone increasing on $h\in[0,J]$ and, hence
\begin{equation}\label{eq:r50}
\frac 12 \tanh (2J/T)\;=\; f_{-1}(J/T,0) \;\le\; \widetilde f(J/T,h/T)\;\le\; f_{-1}(J/T,J/T)\;=\;
\frac{\sinh( 2 J/T) }{1 + \cosh(2J/T ) } 
\end{equation}
Therefore the value of $T^{DC}(J,h)$ lies between $T_{\rm min}$ and $T_{\rm max}$, which are respectively the solutions of the equations
\begin{equation}\label{eq:r51}
\frac 12 \tanh \bigl(2J/T_{\rm min}\bigr)\;=\; \frac14
\end{equation}
and
\begin{equation}\label{eq:r52}
\frac{\sinh( 2 J/T_{\rm max}) }{1 + \cosh(2J/T_{\rm max} ) } \;=\;\frac 14
\end{equation}
This proves inequalities \eqref{Dob1}.

\subsection{Dobrushin-Shlosman type criterion and Proposition \ref{prop:r10}}
\label{sec:DS}

\begin{figure}[h!]
\begin{center}
\subfigure[]{
\includegraphics[scale=0.6]{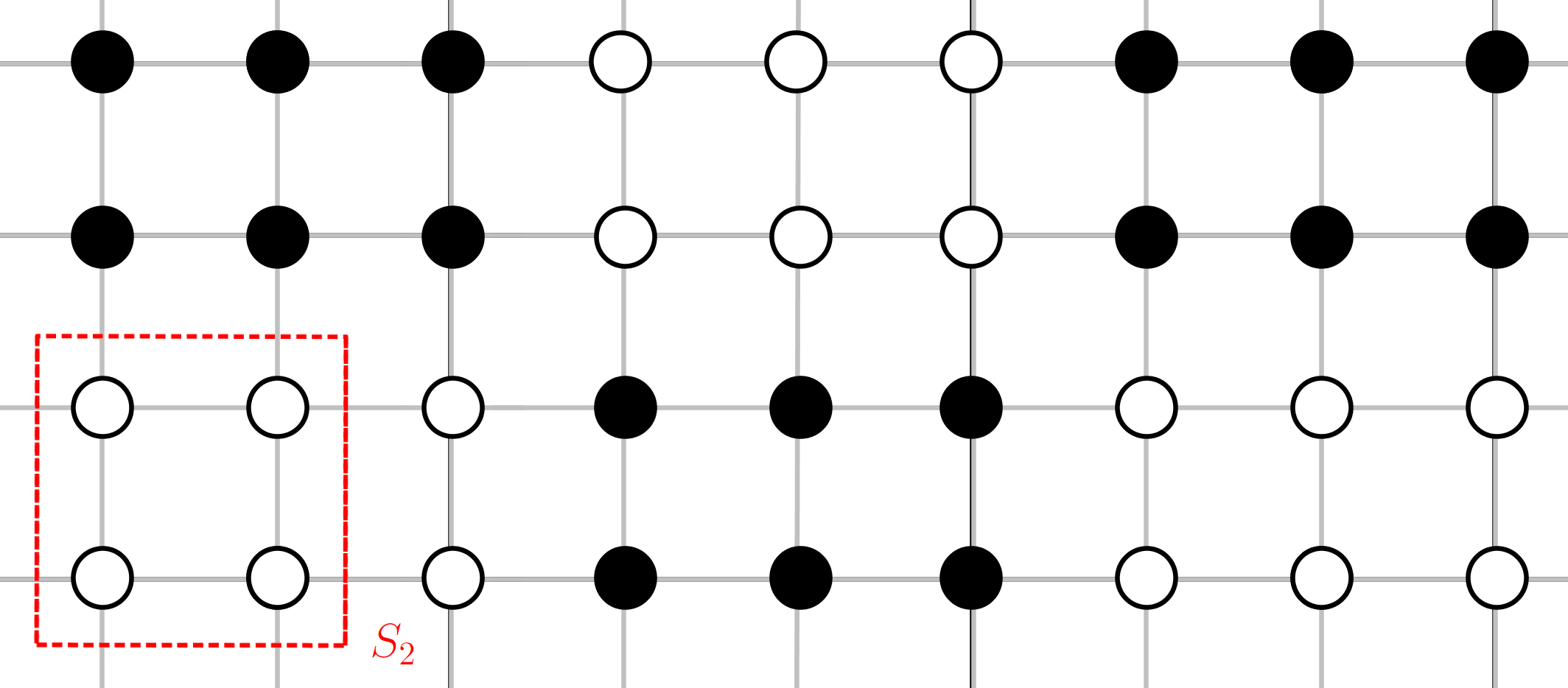}
}
\hspace{1.2cm}
\subfigure[]{
\includegraphics[scale=0.6]{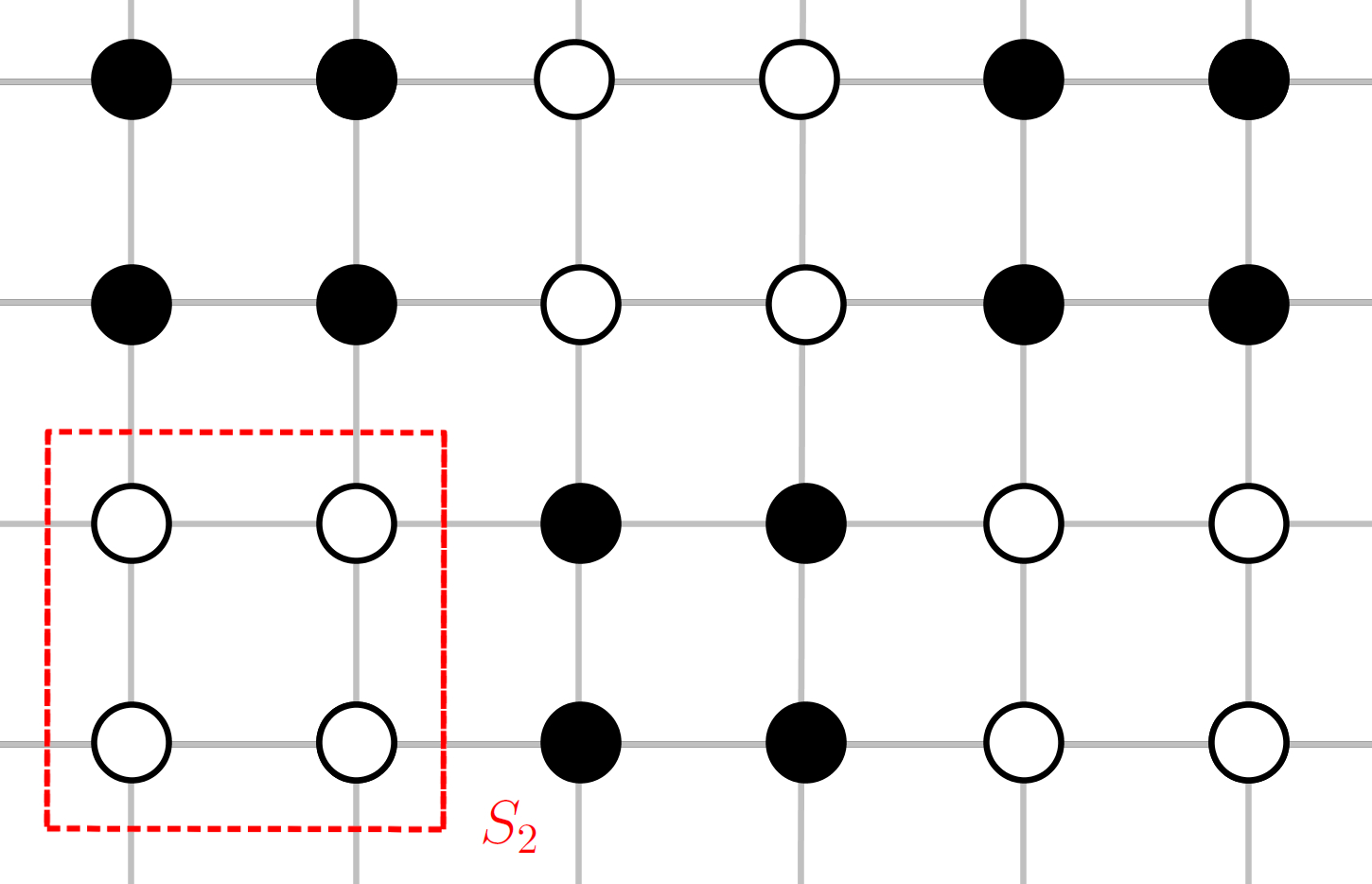}
}
\end{center}\caption{Example for $S_2$, a square of side 2. External fields given by: (a) $L_1=3$ and $L_2=2$. (b) $L_1=L_2=2$.}
\label{figTh}
\end{figure}

We remark that in \eqref{dW} it is not trivial to obtain computational estimates. In Dobrushin et al. \cite{DKS} it was used an optimal coupling instead searching all possibles. In this sense, we introduce a simplified version. As in \cite{AH} we consider $\Lambda$ in the form of square regions $S_n = \{ (i,j): \ 0\le i,j < n \} $ (see Figure \ref{figTh}). We need to consider all configurations of external fields within all possible translations of $S_n$, that is all fields in
\begin{equation}
\label{hsets}
\mathcal{H} (n; L_1,L_2) = \Bigl\{\{h(s)\,,\, s\in S_n + (i,j) \} :  0\le i < L_1, 0\le j < L_2\Bigr \}\;.
\end{equation}
In particular, in the case of Figure \ref{figTh}(b), we have the following configurations of external fields
\begin{eqnarray}
\label{hV2x2}
\mathcal{H} (2; 2,2) =
\left\{ \begin{array}{cccccccccccccccccccccccccccccc}
\circ & \circ & &  \bullet & \bullet &   & \bullet & \bullet &  & \circ & \circ &  & \circ & \bullet  &  &\bullet & \circ &  &\bullet & \circ &  & \circ & \bullet \\
\circ & \circ  &    ,& \bullet & \bullet & , & \circ & \circ & , & \bullet & \bullet & , & \circ & \bullet  & , &\bullet & \circ & , & \circ & \bullet & , &\bullet & \circ \end{array} \right\} \ .
\end{eqnarray}

Therefore, given the periodicity of cell-board external field, we need to state the DS criterion as follows.

\begin{lemma}
\label{DSLem}

Consider a cell-board interaction. The function,

\begin{equation}
\label{alpha}
\alpha_t(n,h,T) = \max_{\mathbf{h} \in \mathcal{H}(n;L_1,L_2)} \displaystyle\sum_{s \in S_n} \sup_{\eta, \eta^{(t)} \in \Omega } \ \Bigl| \mu_{\beta,S_n}^{(\mathbf{h})}(\sigma(s)=+1 | \eta) - \mu_{\beta,S_n}^{(\mathbf{h})}(\sigma(s)=+1 | \eta^{(t)})  \Bigr|,
\end{equation}
where $\mu_{\beta,S_n}^{(\mathbf{h})}$ is the Gibbs-Boltzmann distribution for the given choice of $\mathbf{h}$, satisfies condition \eqref{ThdW} in Theorem \ref{DSthm}.
\end{lemma}

\textbf{Proof}
Let $K(\eta,\eta^{\prime})$, the set of joint measures (couplings) for $\mu_{\beta,S_n}( \cdot | \eta)$ and $\mu_{\beta,S_n}( \cdot | \eta^{\prime})$. For fixed $\mathbf{h} \in \mathcal{H}(n;L_1,L_2)$, $\eta, \eta^{\prime} \in \Omega$, for $\Lambda = S_n$ and the metric \eqref{Thrho}, by definition \eqref{dW} we have that
$$
\begin{aligned}
d_W(\rho_{S_n};\mu_{\beta,S_n}(\cdot| \eta) ,\mu_{\beta,S_n}(\cdot| \eta^{\prime})) & = \inf_{\nu \in K(\eta,\eta^{\prime})}  \sum_{\sigma_{S_n},\sigma^{\prime}_{S_n}} \rho_{S_n} (\sigma_{S_n},\sigma^{\prime}_{S_n}) \nu  (\sigma_{S_n},\sigma^{\prime}_{S_n})  \\
& = \inf_{\nu \in K(\eta,\eta^{\prime})}  \displaystyle\sum_{s \in S_n} \sum_{\sigma_{S_n},\sigma^{\prime}_{S_n}}  \rho_s (\sigma_{S_n}(s),\sigma^{\prime}_{S_n}(s)) \nu  (\sigma_{S_n},\sigma_{S_n}^{\prime})
\end{aligned}
$$
Now, let $$\nu_{s; \eta,\eta^\prime}(\cdot, \cdot) = \inf_{\nu \in K(\eta,\eta^{\prime})} \left\{ \nu( \sigma_{S_n}(s), \sigma_{S_n}^{\prime}(s) ): \sigma_{S_n}(s) \neq \sigma_{S_n}^{\prime}(s) \right\}$$ and define the measure $\nu^* (\cdot, \cdot):=\prod_{s\in S_n} \nu_{s; \eta,\eta^\prime}(\cdot, \cdot)$. In addition, as a metric $\rho_t$ in \eqref{Thrho}, we use the discrete metric
\begin{equation*}
\rho_s(\sigma(s),\eta(s))=
\left\{
\begin{array}{rl}
1,&\text{ if }\sigma(s)\neq\eta(s),\\
0,&\text{ if }\sigma(s) = \eta(s),
\end{array}
\right.
\end{equation*}
where $\sigma(s),\eta(s) \in  \{-1,+1\}$. Then
\begin{equation*}
\begin{array}{l}
d_W(\rho_{S_n};\mu_{S_n}(\cdot| \eta) ,\mu_{S_n}(\cdot| \eta^{\prime})) \le  \displaystyle\sum_{s \in S_n} \sum_{\sigma_{S_n},\sigma^{\prime}_{S_n}}  \rho_s (\sigma_{S_n}(s),\sigma^{\prime}_{S_n}(s)) \nu^*  (\sigma_{S_n},\sigma_{S_n}^{\prime})
\\[0.5cm] \le  \displaystyle\sum_{s \in S_n}  \left( \nu_{s; \eta,\eta^\prime}(\sigma_{S_n}(s)=+1, \sigma_{S_n}^{\prime}(s) =-1 ) + \nu_{s; \eta,\eta^\prime}(\sigma_{S_n}(s)=-1, \sigma_{S_n}^{\prime}(s) =+1) \right)
\\[0.5cm] \le \alpha_t(n,h,T).
\end{array}
\end{equation*}

\textbf{Remark:} The maximum in \eqref{alpha} is given by all translations of external fields as stated in \eqref{hsets}.
In this sense, we define the sequence of parameters
\begin{equation}
\label{DSgamma}
\gamma_n(J/T,h/T,L_1,L_2) : =  
\frac{1}{n^2} \displaystyle\sum_{t \in \partial S_n} \alpha_{t}(n,h,T),
\end{equation}
for the Proposition \ref{prop:r10}.

Essentially, we are using total variation \eqref{dTV} in the constants $\alpha_{t}$, although this arbitrary choice, the estimates obtained by DS criterion are better than DP and DC.

In this sense, the version of DS criterion adopted here is, therefore, determined by the constants \eqref{DSgamma}. These are the functions defining the DS lines \eqref{hDS1} and numerically computed in Figures \ref{fig:Anti}, \ref{fig:V2x2} and \ref{fig:L2x1}.

\smallskip

Finally, let consider \eqref{DSgamma}, note that $\gamma_n$ depend on $\mathcal{H}(n;L_1,L_2)$ as defined in \eqref{hsets}, the identities \eqref{DS1}-\eqref{eq:r15} follow from the following arguments. First, \eqref{DS1} from the fact that $\mathcal{H} (n;L_1,L_2) = \mathcal{H} (n;n,n)$ for all $L_1$ and $L_2$ such that $\min\{L_1,L_2\} \ge n$, which implies that the maximum in \eqref{DSgamma} is the same for each $L_1$x$L_2$ cell. Inequality \eqref{DS2} is a consequence of the fact that $\mathcal{H}(n;n-1,n-1) \subseteq \mathcal{H}(n;n,n)$.  
\smallskip

The proof of \eqref{eq:r15} relies on the following observations:
\begin{itemize}
\item By \eqref{DS1}, $l^{DS}(2;J,T,L_1,L_2) =l^{DS}(2;J,T,2,2)$ for all $L_1, L_2 \ge 2$.  
\item $\mathcal{H}(2;2,2) \supseteq \mathcal{H}(2;1,2) , \mathcal{H}(2;2,1)  \supseteq  \mathcal{H}(2;1,1)$.  
\item $l^{DS}(2;J,T,2,2) =l^{DS}(2;J,T,1,1)$ (this can be checked with a symbolic computing program).
\end{itemize}

\medskip

\textbf{Acknowledgments}
We thank anonymous referee for the valuable contributions. MGN was supported by Funda\c{c}\~ao de Amparo \`a Pesquisa do Estado de S\~ao Paulo (FAPESP) grant 2015/02801-6, and Fondecyt Iniciaci\'on 11200500. He thanks L.R. Fontes and E. Presutti for their help in that project, also kind hospitality of Gran Sasso Science Institute and Institute for Information Transmission Problems (IITP). The researches of E. Pechersky was carried out at the IITP Russian Academy of Science. A. Yambartsev thanks Conselho Nacional de Desenvolvimento Cient\'ifico e Tecnol\^ogico (CNPq) grant 301050/2016-3 and FAPESP grant 2017/10555-0.

\appendix
\section{Numerical algorithms}
\label{sec:proof}

We present the codes used to obtain the numerical results in Section \ref{sec:model}. Remember that we are interested in the region $h< 4J$, that is, we consider $h<4$ for all numerical calculations. In particular, we will check DP and Dobrushin conditions on the discretized values $T\in \mathcal{T}$ and $h \in \mathcal{H}$, where
\begin{equation}
\label{TH}
\mathcal{T}= \{t=0.001 \times k \ | \ k=0,1, \ldots, 7000\} \ \text{ and } \ \ \mathcal{H}=\{h=0.1\times k \ | \ k=0,1, \ldots, 45\}.
\end{equation}

\bigskip
\underline{Disagreement percolation}
\bigskip

\noindent
We just noted that the disagreement percolation condition does not depend on values of cell sizes: for any $L_1$ and $L_2$ the temperature defined by (\ref{T0DP}), $T^{DP}(h)\equiv T^{DP}(1,h,L_1,L_2)$, is the same. For the condition (\ref{uniccond1}) we use the result from \cite{vdBE}, that provide the following bound for critical percolation probability: $p_c (\Z^2) > 0.556$. Thus, for each $h \in \mathcal{H}$ we numerically find the root $T^{DP}(h)$ of the equation inside \eqref{T0DP}, where the left-hand function was defined by \eqref{11}. In order to obtain the DP estimates for uniqueness region
we run the (simple) Algorithm \ref{DPalgorithm}.

\bigskip
 
\begin{algorithm}
\caption{Disagreement Percolation condition}\label{DPalgorithm}
\begin{algorithmic}[1]
\FORALL{ $h\in \mathcal{H}$} 
\STATE find numerically the root $t(h)$ of equation: $p(h,t(h))-0.556=0$ 
\STATE put the found root to array $T$
\ENDFOR
\STATE plot values of arrays $(\mathcal{H}, T)$
\end{algorithmic}
\end{algorithm}

\bigskip
\underline{Dobrushin condition}
\bigskip

\noindent

Analogously for Dobrushin criterion, for each $h \in \mathcal{H}$, we obtain $T^{DC} (1,h)$ defined by \eqref{T0DC}. 
Moreover, in the proof of Theorem~\ref{PropoDo} the function $T^{DC} (1,h)$ was defined as an implicit function by formulas \eqref{DCcritfunc1} and \eqref{DCcritfunc2}. Thus the algorithm which we run in order to construct a bound for uniqueness region is simple again and is given in Algorithm \ref{DCalgorithm}.

\begin{algorithm}
\caption{Dobrushin condition}\label{DCalgorithm}
\begin{algorithmic}[1]
\FORALL{ $h\in \mathcal{H}$} 
\IF{ $h\in [0,2]$ }
\STATE find numerically the root $t(h):$ of the equation $f_{-1}(h)-0.25=0$ (see \eqref{DCcritfunc1},\eqref{DCcritfunc2})
\STATE put the found root to array $T$
\ENDIF
\IF{ $h\in [2,4]$ }
\STATE find numerically the root $t(h):$ of the equation $f_{-3}(h)-0.25=0$ (see \eqref{DCcritfunc1},\eqref{DCcritfunc2})
\STATE put the found root to array $T$
\ENDIF
\ENDFOR
\STATE plot values of arrays $(\mathcal{H}, T)$
\end{algorithmic}
\end{algorithm}

\bigskip

\underline{Dobrushin-Shlosman criterion}
\bigskip

\noindent
It is the computationally time consuming case. We use
\begin{equation}
\label{THDS}
\mathcal{T}= \{t=0.002 \times k \ | \ k=1, \ldots, 205\} \ \text{ and } \ \ \mathcal{H}=\{h=0.05\times k \ | \ k=1, \ldots, 80\}.
\end{equation}
The numerical implementation of \eqref{hDS1} will be the following: for fixed $L_1$ and $L_2$, given $n\in\mathbb{N}$, at each $T \in \mathcal{T}$ we calculate
\begin{equation*}
h^{DS}(n;1,T,L_1,L_2)=\min \{ h \in \mathcal{H}: \text{ condition } \ \eqref{DSgamma} \text{ holds} \}.
\end{equation*}

This curve is enough in most of the cases. However, in the curve $l^{DS}(3;1,\infty,2)$ in Figure \ref{fig:L2x1}, it was necessary to detail the analysis and complement the estimate $h^{DS}$ with the curve
\begin{equation*}
T^{DS}(n;1,h,L_1,L_2)=\min \{ T \in \mathcal{T}: \text{ condition } \ \eqref{DSgamma} \text{ holds} \},
\end{equation*}
for each $h \in \mathcal{H}$.
\medskip

In Algorithm \ref{DSalgorithm} we introduce the pseudo-code used to obtain all the numerical results with DS criterion. In the case of $T^{DS}$ the algorithm is analogous.


\bigskip

\begin{algorithm}
\caption{Dobrushin-Shlosman condition}\label{DSalgorithm}
\begin{algorithmic}[1]
\STATE fix parameters $n, L_1, L_2$
\FORALL{ $t \in \mathcal{T}$} 
\STATE $h=0$ the initial value of external field 
\STATE $Ind=0$ the indicator of the DS condition
\WHILE{ $Ind =0$}
\FORALL{ $\mathbf{h} \in \mathcal{H}(n;L_1,L_2)$ }
  \STATE $Sum=0$, the sum in the condition \eqref{DSgamma}
  \FORALL{ $s \in S_n$ and $t \in  \partial S_n$ }
  \STATE calculate $\alpha_{st}(\mathbf{h})$, see \eqref{alpha}
  \STATE $Sum = Sum + \alpha_{st}(\mathbf{h})$
  \ENDFOR
\STATE put $Sum/n^2$ in array $\Gamma$
\ENDFOR
\IF{ $\max (\Gamma) < 1$ }
\STATE $Ind=1$
\STATE put the current value of $h$ in array $\mathcal{H}$
\ELSE 
\STATE $h=h+0.05$
\ENDIF
\ENDWHILE
\ENDFOR
\STATE plot values of arrays $(\mathcal{H}, \mathcal{T})$
\end{algorithmic}
\end{algorithm}

\bigskip

{\scriptsize

{\sc New York University Shanghai, 1555 Century Avenue, Pudong, Shanghai, China.} E-mail address: rf87@nyu.edu\\

{\sc Departamento de Estad\'i{}stica, Universidad del B\'io-B\'io. Avda. Collao 1202, CP 4051381, Concepci\'on, Chile.} E-mail address: magonzalez@ubiobio.cl\\

{\sc Institute for Information Transmission Problems, 19, Bolshoj Karetny per., Moscow, Russia.} E-mail address: pech@iitp.ru\\

{\sc Instituto de Matem\'atica e Estat\'i{}stica, Universidade de S\~ao Paulo. Rua do Mat\~{a}o, 1010, CEP 05508-090, S\~{a}o Paulo, SP, Brazil.} E-mail address: yambar@ime.usp.br\\
}

\end{document}